\documentclass[a4paper,14pt]{extarticle}

\usepackage[T2A,T1]{fontenc}
\usepackage[english]{babel}

\usepackage{graphicx}
\usepackage{subcaption}
\usepackage{indentfirst}

\usepackage{hyperref}
\usepackage{tabularx}
\usepackage{array}
\usepackage{multirow}

\usepackage{gensymb}
\usepackage{amsmath}
\usepackage{bm}

\usepackage{cite}

\title{Testing Docker Performance for HPC Applications}
\author{
  Alexey Ermakov\      \texttt{ermakov.as@mipt.ru}
  \and
  Alexey Vasyukov\      \texttt{a.vasyukov@phystech.edu}
}
\providecommand{\keywords}[1]{\textbf{\textit{Keywords:}} #1}

\newcommand\fundnote[1]{%
  \begingroup
  \renewcommand\thefootnote{}\footnote{#1}%
  \addtocounter{footnote}{-1}%
  \endgroup
}

\begin{document}

\maketitle

The main goal for this article is to compare performance penalties
when using KVM virtualization and Docker containers for creating iso-
lated environments for HPC applications. The article provides both data
obtained using synthetic tests (High Performance
Linpack) and real life applications (OpenFOAM). The article highlights
the influence on performance of major infrastructure configuration options
 -- CPU type presented to VM, networking connection type used.


\fundnote{The research was supported by RFBR grant 15-29-07096} 

    \section{Introduction}

    One of the most important issues related to high performance computing that
    one may encounter is the availability of certain execution environment. It
    means that many scientific programs require a specific set of dependencies
    (such as compilers, runtime libraries etc.), that often may even conflict
    with dependencies of other software. There is a number of ways to solve the
    issue, one the most mature technologies that is used for such a purpose is
    a virtualization. Despite tha fact that virtualization provides full
    environment isolation, by-design it has some performance penalty. Another
    approach to provide isolated environment is operating-system-level
    virtualization that implies all such environments have common kernel and
    separate isolated user-space libraries. The main goal for this article is to
    compare performance penalties when using two mentioned ways of creating
    isolated environment (KVM and Docker containers, to be precise).

    \section{Related work}
    Cloud computing environments for HPC applications are commonly based on KVM
    for virtualization and isolation and OpenStack for cluster management, 
    auto-provision and user self-service. An example of the system
    based on these technologies can be found in \cite{zih2015}, describing an 
    experience of Technische Universitat Dresden. Similar KVM-based clusters
    are deployed in different organizations over the world. However, performance
    penalties for real life applications may be significant when running in
    virtualized environment \cite{ibm2014}. Container-based systems for HPC
    applications emerge during recent years \cite{lbl2015, nersc2015, arXiv-1509-08231} 
    and benchmarks look promising \cite{qnib2014, arXiv-1601-03872}. This article also
    contributes to public benchmarks of KVM and Docker containers for HPC applications.

    \section{Virtual machines and containers}

\begin{figure}
        \begin{center}

            \includegraphics[width=\textwidth]{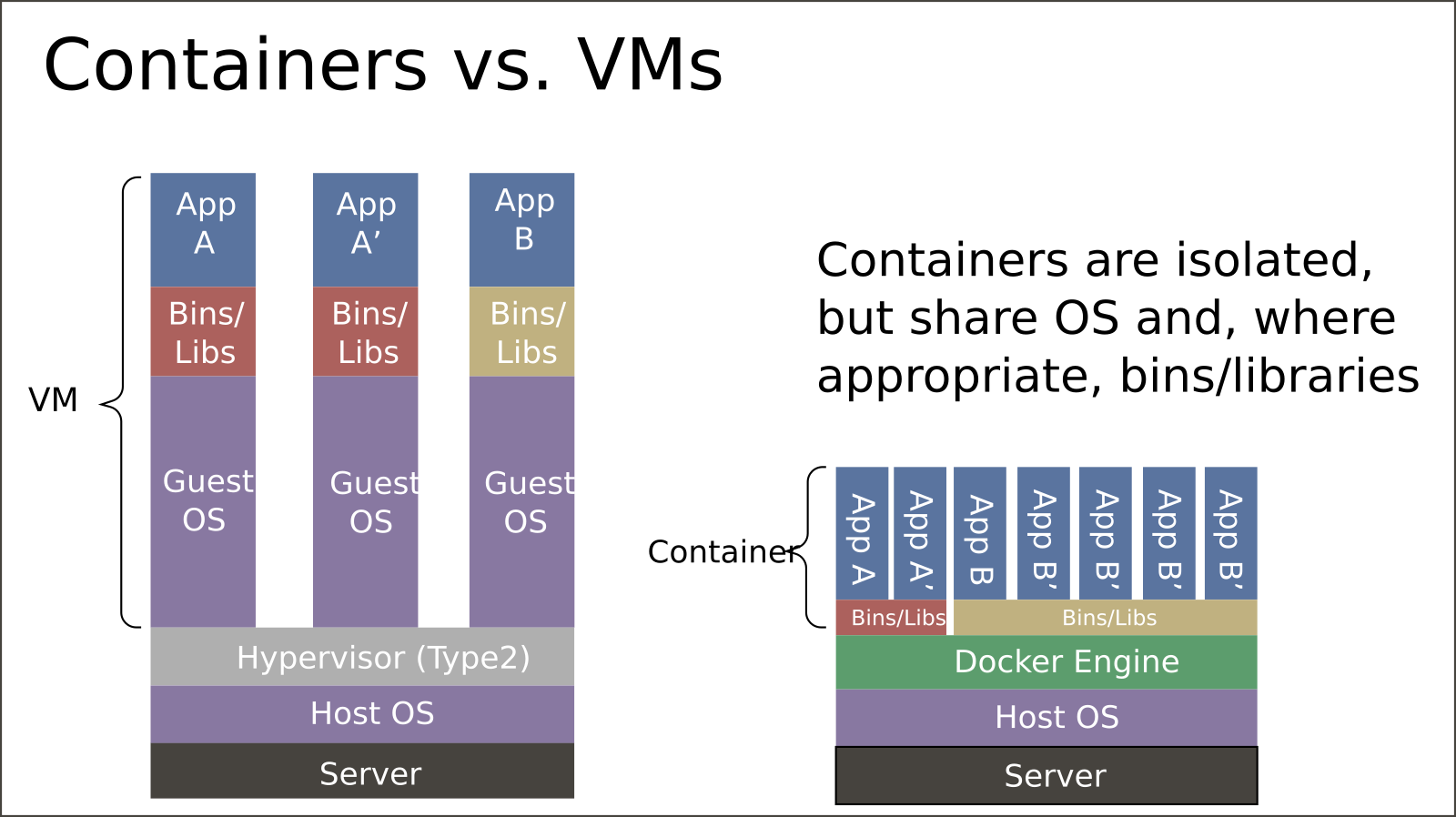}
            \caption{Comparsion of virtualization and containers architecture\protect\footnotemark}
            \label{pic:architecture-comparsion}
        \end{center}
    \end{figure}

    As it was mentioned earlier, the main difference (see fig. \ref{pic:architecture-comparsion})
    between virtualization and containerization is that containers share the
    same kernel and maybe even some host devices, when each virtual machine has
    its own kernel and virtualized devices (e.g. network card)\footnote{In this
    article we do not consider usage of paravirtualization or any
    <<passthrough>> technologies to make host devices available to virtual
    machine}. More information on used technologies may be found in official
    documentation for KVM\cite{kvm, qemu} and Docker\cite{docker}.

    \footnotetext{Image courtesy of Linux Magazine \url{http://www.linux-magazine.com/Issues/2015/171/Docker\#article_f2}}

    \section{Benchmark setup and methodics}

\begin{figure}[h!]
        \begin{center}
            \includegraphics[width=0.5\textwidth]{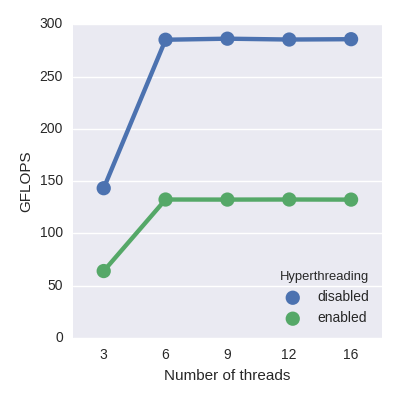}
            \caption{Hyperthreading performance impact according to Intel
            Linpack benchmark results}
            \label{pic:ht}
        \end{center}
    \end{figure}

    To perform benchmark the following setup was used: two identical hosts with
    Intel Core i7-5820K CPU (6 physical cores) and 64 GB RAM, connected with QDR
    Infiniband and 100 MB/s Ethernet. Hyperthreading was disabled using
    corresponding BIOS settings, since it drastically decreases performance
    (see fig. \ref{pic:ht}). MPICH was used as MPI implementation, because it's
    a bit faster than OpenMPI and does not require any configuration to execute
    program on two hosts when they belong to different subnets (this is very
    important to be able to inter-connect virtual machines and containers.)

    Benchmarks described in this article use Intel Linpack benchmark\cite{intel-linpack},
    High-performance linpack\cite{hpl} and interFoam solver
    as <<real world>> application from OpenFOAM\cite{openfoam}. All experiments were run 10 times
    to reduce statistical errors, so each plot shows mean value for measurement
    and error bars for confidence interval of 0.95.

    \begin{figure}[h!]
        \centering
        \subcaptionbox{QDR infiniband connection}{
            \includegraphics[width=0.45\linewidth]{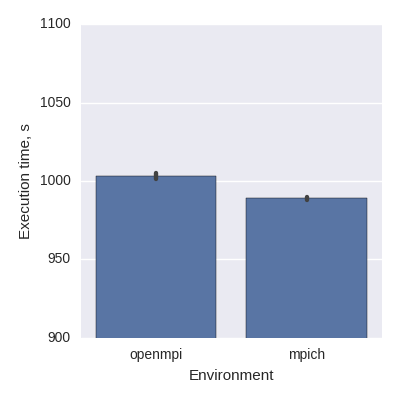}
        }
        \subcaptionbox{100 Mb/s Ethernet connection}{
            \includegraphics[width=0.45\linewidth]{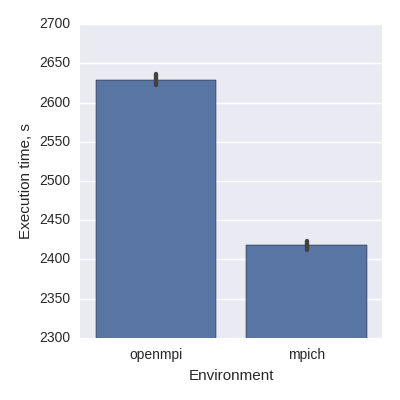}
        }
        \caption{Comparsion of OpenMPI and MPICH performance using HPL}
        \label{pic:mpich-vs-openmpi}
    \end{figure}

    \section{Benchmark}

    \subsection{1 host benchmarks}

    First of all let's see how usage of containers and virtualization impacts
    performance. The following tests were performed on a single to host to
    avoid networking influence on performance results.

    \begin{figure}[h!]
        \begin{center}
            \includegraphics[width=0.5\textwidth]{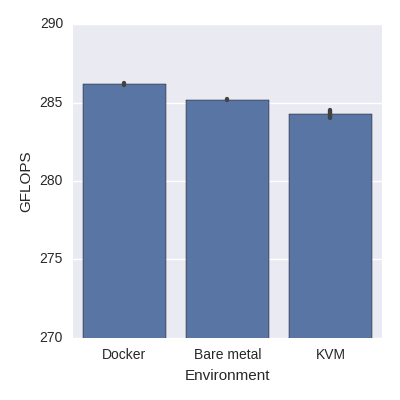}
            \caption{Performance comparsion of KVM and Docker on a single host
            using Intel Linpack}
            \label{pic:bm-docker-kvm}
        \end{center}
    \end{figure}

    Results shown on a fig. \ref{pic:bm-docker-kvm} should be treated as
    follows: there is  is no significant difference in performance when running
    CPU-intensive highly-optimized application in KVM, Docker or on bare metal.
    It should be noted that in these tests Intel Linpack demonstrates ~90\% of
    theoretical CPU performance, thus performance comparsion may be considered
    reliable. We can see that Docker shows a bit better performance even than
    bare metal, but it should be considered as statistical error. Another cause
    for this may be operating system scheduler that for some reason gives a bit
    more priority for containerized processes.

    In previos test QEMU was run with host-model CPU set, that's why it showed
    pretty good performance. In case of different CPU type there is a huge
    performance spread: depending on exact CPU model used to run a virtual
    machine KVM may be up to 5 times slower than a bare metal.

    \begin{figure}[h!]
        \begin{center}
            \includegraphics[width=0.5\textwidth]{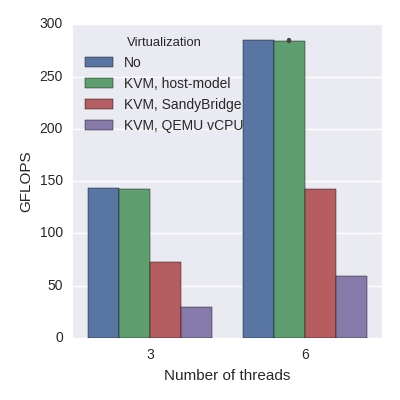}
            \caption{KVM performance spread}
            \label{pic:linpack-kvm}
        \end{center}
    \end{figure}

    \subsection{2 hosts benchmarks}

    Another component, besides CPU, that has a major performance influence is
    networking. The next series of tests demonstrate what performance impact one
    may have when using unappropriate networking stack.

\begin{figure}[h!]
        \begin{center}
            \includegraphics[width=0.5\textwidth]{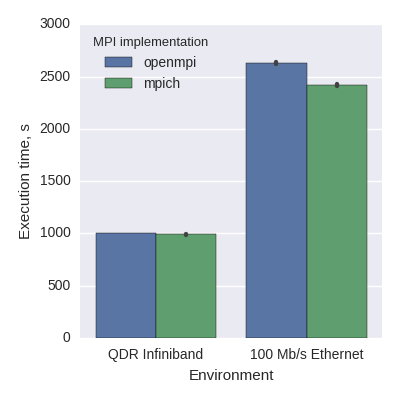}
            \caption{HPL performance comparsion using different
            inter-connection}
            \label{pic:hpl-mpich-openmpi}
        \end{center}
    \end{figure}

    First thing to note is quite obvious but anyway should be mentioned:
    networking type matters. According to fig. \ref{pic:hpl-mpich-openmpi} HPL
    performs about 2.5 slower on 100 Mb/s Ethernet than on QDR
    Infiniband.

    \begin{figure}[h!]
        \begin{minipage}[c]{0.5\textwidth}
            \begin{center}
                \includegraphics[width=\textwidth]{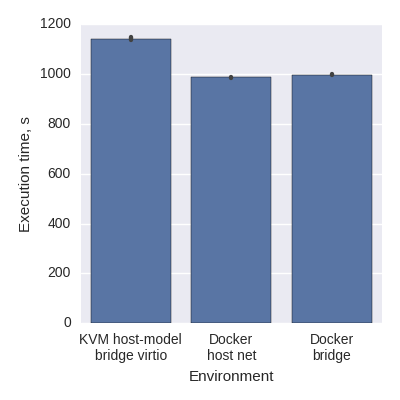}
                \caption{Docker and KVM networking performance comparsion}
                \label{pic:hpl-docker-vs-kvm}
            \end{center}
        \end{minipage}
        \begin{minipage}[c]{0.5\textwidth}
            \begin{center}
                \includegraphics[width=\textwidth]{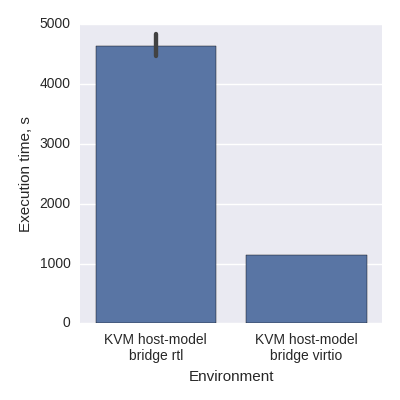}
                \caption{virtio and rtl networking performance comparsion}
                \label{pic:hpl-kvm-rtl-vs-virtio.png}
            \end{center}
        \end{minipage}
    \end{figure}

    Another thing that should be taken into account is the way virtual machine
    or container is connected to network. For HPL tests there is almost no
    difference (see fig. \ref{pic:hpl-docker-vs-kvm}) in performance of
    dockerized network application between bridged connection and host
    networking stack, KVM virtio performs about 15\% slower while KVM rtl is
    almost 5 times slower (see fig. \ref {pic:hpl-kvm-rtl-vs-virtio.png}) than
    host netwoking stack.

\begin{figure}[h!]
        \begin{center}
            \includegraphics[width=0.5\textwidth]{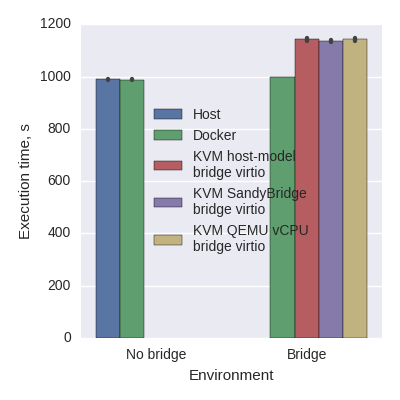}
            \caption{CPU type influence on performance of HPL}
            \label{pic:hpl-host-vs-docker-vs-kvm}
        \end{center}
    \end{figure}

    Earlier we've noticed that CPU type used to run virtual machine has
    significant influence on overall performance. Let's see if it still
    applies to distributed MPI application. According to pic.
    \ref{pic:hpl-host-vs-docker-vs-kvm} CPU used to run virtual machine does not
    affect resulting performance. Reason for this behaviour is the fact that HPL
    is not as CPU-intensive as Intel Linpack: network performance is more
    important for HPL rather than CPU performance.

\begin{figure}[h!]
        \begin{minipage}[c]{0.5\textwidth}
            \begin{center}
                \includegraphics[width=\textwidth]{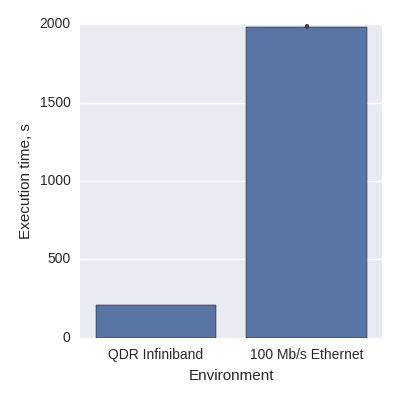}
                \caption{Networking type influence}
                \label{pic:openfoam-ib-vs-eth}
            \end{center}
        \end{minipage}
        \begin{minipage}[c]{0.5\textwidth}
            \begin{center}
                \includegraphics[width=\textwidth]{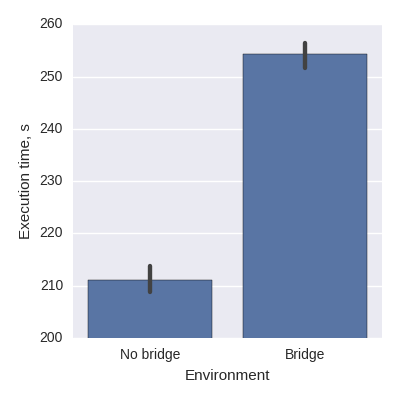}
                \caption{Network connection type influence}
                \label{pic:openfoam-docker-host-vs-bridge}
        \end{center}
        \end{minipage}
    \end{figure}

\begin{figure}[h!]
        \begin{center}
            \includegraphics[width=0.8\textwidth]{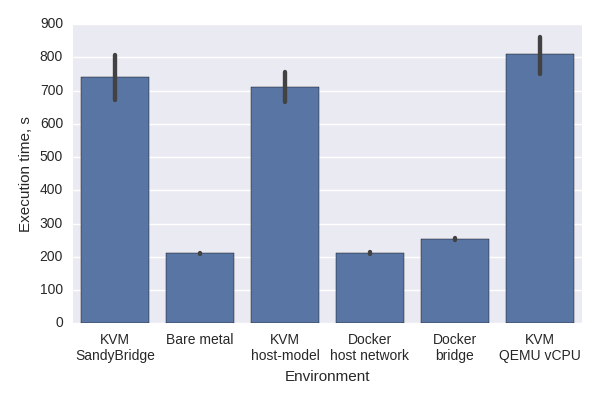}
            \caption{Comparsion of interFoam performance depending on used environment}
            \label{pic:openfoam-bare-metal-vs-docker-vs-kvm}
        \end{center}
    \end{figure}

    Finally, let's see how <<non-synthetic>> distributed MPI application
    performs depending on used environment. We used interFoam solver from
    OpenFOAM in few scenarios to realize how virtualization type and kind
    on networking connection influences overall performance. Results are shown
    on figs. \ref{pic:openfoam-ib-vs-eth} -- \ref{pic:openfoam-bare-metal-vs-docker-vs-kvm}.
    As we can see interFoam performs almost 10 times slower on 100 Mb/s Ethernet
    rather than on QDR Infiniband, bridge performance impact is about 20\%.

    Most important results are shown on fig. \ref{pic:openfoam-bare-metal-vs-docker-vs-kvm}:
    a real <<non-synthetic>> application is more than 2 times slower in KVM
    with virtio networking than the same application in Docker with host or
    bridged networking.

    \section{Conclusion}

    Performance of virtualized or containerized applications depends on many
    factors, such as CPU type (for virtualization) and networking type.
    In some cases performance may degrade up to 10 times, thus environment to
    run application in must be carefully selected and verified. A really
    important thing is that <<synthetic>> benchmark does not provide you with a
    full exhaustive information necessary to decide if environment is suitable
    for an application or not. That's why it's strongly recommended to run
    benchmarking on exact application you're going to run when considering
    virtualization or containerization as an option for HPC.

    \bibliographystyle{hunsrt}
    \bibliography{docker_kvm_bare_metal_performance}

\end{document}